\documentclass[FBSedit,FBSmath,ecsub]{FBSsuppl}
\usepackage{amsfonts}
\usepackage{amssymb}
\usepackage{epsfig}
\newcommand{\ga}{\alpha}
\newcommand{\gb}{\beta}
\newcommand{\gc}{\gamma}
\newcommand{\gd}{\delta}

\newcommand{\gl}{\lambda}
\newcommand{\gs}{\sigma}
\newcommand{\gS}{\Sigma}
\title{Clusters and condensates in Fermi systems}
\author{Michael Beyer\thanks{\textit{E-mail address:} 
michael.beyer@physik.uni-rostock.de}}
\institute{Fachbereich Physik, Universit\"at Rostock, 18051 Rostock, Germany}

\begin{document}

\maketitle
\begin{abstract}
  Superconductivity, superfluidity, condensation, cluster formation,
  etc. are phenomena that might occur in many-particle systems. These
  are due to residual interactions between the particles. To explain
  these phenomena consistently in a microscopic approach, at some
  point, one needs to solve few-body equations that are modified
  because of the Pauli principle and the interactions of the many
  particles around.
\end{abstract}

\section{Introduction}
Correlations in many-particle systems are responsible for a number of
interesting and exciting properties of the system. In an infinite,
equilibrated system the presentation of these properties (or phases of
matter) are usually organized in a phase diagram. This is a plot of
the dominant state of matter separated by ``critical lines'' depending
on the two basic variational parameters of quantum statistics relevant
for a grand canonical ensemble, i.e. the temperature $T$ and the
chemical potential $\mu$. Instead of the chemical potential $\mu$ that
is difficult to access experimentally one might by use of proper
thermodynamic relations chose the particle density $n(\mu,T)$.
Note, however, that the relation between $\mu$ and $n(\mu,T)$ is not
simple and depends on the specific system considered.

\begin{figure}[t]
\begin{center}
\epsfig{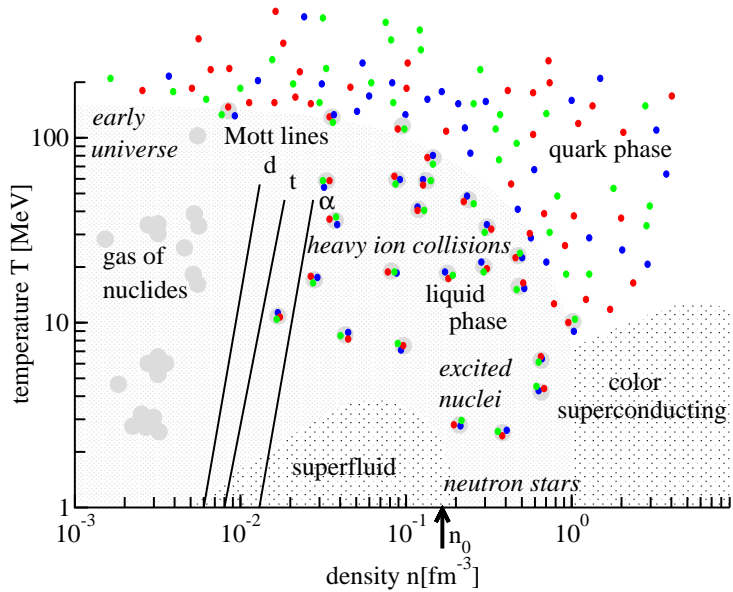}
\caption{\label{fig:phase} 
  Schematic view of the phase diagram of nuclear matter. The phase
  diagram is empirical accessible by heavy ion collisions, excited
  nuclei, observation of neutron stars and the early universe as
  indicated in the diagram. New plans at GSI aim at exploring the
  color superconducting phase as well.}
\end{center}
\end{figure}

The phase diagram becomes particularly rich when the chemical
potential $\mu$ is in the order of MeV to GeV. This is at rather large
particle densities, where nuclei are close together. This is the
domain where Quantum Chromodynamics (QCD) dominates the dynamics of
the system.  Since nucleons are made of quarks it is obvious that the
relevant degrees of freedom to describe the system might change from
nucleons (nuclear matter) to quarks (quark matter) at some point. In
addition gluons might appear as additional degrees of freedom that are
otherwise absent at lower densities. Because of the free (color)
charge this phase is called a (quark gluon) plasma. Indeed, lattice
QCD and model calculations suggest a variety of phenomena in the QCD
dominated phase of matter.  These are superfluidity, critical points,
first and second order phase transitions, color super conductivity,
plasma phase, among others. A sketch of the phase diagram is given in
Fig.~\ref{fig:phase}. To explain all these different states of matter
one needs a treatment that goes beyond the simple quasiparticle
picture.

To tackle the complicated many-particle system the Green function
approach is a good starting point~\cite{fet71}.  Equations for Green
functions are hierarchally coupled and hence an approximate solution
of the problem is unavoidable. If we are interested in two-body
correlations only, the three-body Green functions are approximated.
Quite a few phenomena are already accessible at this level. However,
there are good reasons to go beyond two-particle correlations, e.g.:
\begin{itemize}
\itemsep=0pt
\item Particle production even in a dense environment such as deuteron
  formation in a heavy ion reaction, need a third particle to conserve
  energy-momentum~\cite{Kuhrts:2001zs}.
\item To study the properties of
  $\alpha$-particles~\cite{Beyer:2000ds} or determine the critical
  temperature of a possible $\alpha$-particle
  condensate~\cite{alpha,Tohsaki:2001an,Suzuki:2002me}
  needs an in-medium four-body equation.
\item Recent results in the Hubbard model indicate, that
  three-particle contributions may lead to a different (lo\-wer)
  critical temperature compared to the simple Thouless
  criterion~\cite{letz}. Question of this type have not been addressed
  for nuclear matter.
\item The chiral phase transition is often discussed along with a
  confinement-deconfinement transition based on investigating mesons
  (quark-antiquark states). Does this transition happen for nucleons
  (three-quark states) at the same density/temperature?
\end{itemize} 
Some of these issues related to the nuclear matter phase are addressed
in the present paper. The ones related to quark matter are given by
Mattiellio et al~\cite{mat2002}.

\section{Theory}

We use Dyson equations to tackle the many-particle problem, see e.g.
Ref.~\cite{duk98}. This enables us to decouple the hierarchy of Green
functions. The Dyson equation approach used here is based on two
ingredients: {\it i)} all particles of a cluster are taken at equal
time {\it ii)} the ensemble averaging for a cluster is done for an
uncorrelated medium.  The resulting decoupled Green functions may be
economically written as resolvents in the $n$-body space, where
$n=2,3,4,\dots$ is the number of particles in the considered cluster.

The solution of the one-particle problem in Hartree-Fock approximation
leads to the following quasi-particle energy
\begin{equation}
\varepsilon_1 = \frac{k^2_1}{2m_1}+\sum_{2}V_2(12,\widetilde{12})f_2
\simeq \frac{k^2_1}{2m_1^{\rm eff}}+\gS^{\rm HF}(0).
\label{eqn:quasi}
\end{equation}
The last equation introduces the effective mass that is a valid
concept for the rather low densities considered here and $\mu^{\rm
  eff}\equiv \mu - \gS^{\rm HF}(0)$.  The Fermi function $f_i\equiv
f(\varepsilon_i)$ for the $i$-th particle is given by
\begin{equation}
f(\varepsilon_i) = \frac{1}{e^{\gb(\varepsilon_i - \mu)}+1}.
\end{equation}
The resolvent $G_0$ for $n$ noninteracting quasiparticles is
\begin{equation}
G_0(z) = (z-  H_0)^{-1}
N \equiv R_0(z) N,\qquad H_0 = \sum_{i=1}^n \varepsilon_i
\end{equation}
where $G_0$, $H_0$, and $N$ are formally matrices in $n$ particle
space.  The Matsubara frequency $z_\gl$ has been analytically
continued into the complex plane, $ z_\gl\rightarrow
z$~\cite{fet71}. The Pauli-blocking for $n$-particles is
\begin{equation}
N=\bar f_1\bar f_2 \dots \bar f_n
\pm f_1f_2\dots f_n,\qquad\bar f=1-f
\end{equation}
where the upper sign is for Fermi-type and the lower for Bose type
clusters. The full
resolvent $G(z)$ is given by
\begin{equation}
G(z)=(z-H_0-V)^{-1}{N}
, \qquad
V\equiv \sum_{\mathrm{pairs}\;\ga} N_2^{\ga}V_2^{\ga}.
\end{equation}
Note that $V^\dagger\neq V$.
For the two-body case as well as for a two-body subsystem embedded in
the $n$-body cluster the standard definition of the $t$ matrix leads
to the Feynman-Galitskii equation for finite temperature and
densities~\cite{fet71},
\begin{equation}
T_2^\ga(z) =   V_2^\ga + 
 V_2^\ga  N^\ga_2 R_0(z)  T_2^\ga(z).
\label{eqn:T2}
\end{equation}
Introducing the  Alt Grassberger Sandhas (AGS)
transition operator $U_{\alpha\beta}(z)$
the effective inhomogeneous in-medium AGS equation reads
\begin{equation}
U_{\alpha\beta}(z)= (1-{\delta}_{\alpha\beta})R^{-1}_0(z)+
\sum_{\gamma\neq \alpha}
{N^\gc_2}
T_2^\gamma(z)R_0(z)
U_{\gamma\beta}(z).
\label{eqn:T3}
\end{equation}
The homogeneous in-medium AGS equation uses the form factors defined by
\begin{equation}
|F_\gb\rangle\equiv\sum_\gc\bar\gd_{\gb\gc} { N_2^\gc}  V_2^\gc 
|\psi_{B_3}\rangle
\end{equation}
to calculate the bound state $\psi_{B_3}$
\begin{eqnarray}
|F_\ga\rangle
&=&\sum_\gb \bar\gd_{\ga\gb}  
{N_2^\gb} T_2^\gb(B_3) R_{0}(B_3)|F_\gb\rangle.
\end{eqnarray}
Finally, the four-body bound state is described by
\begin{equation}
|{\cal F}^\gs_\gb\rangle=\sum_{\tau\gc} \bar\gd_{\gs\tau}
U^\tau_{\gb\gc}(B_4)  R_0(B_4) { N_2^\gc} 
T_2^\gc(B_4) R_0(B_4) |{\cal F}^\tau_\gc\rangle,
\end{equation}
where $\ga\subset\gs,\gc\subset\tau$ and $\gs,\tau$ denote the
four-body partitions. The two-body input is given in (\ref{eqn:T2})
and the three-body input by (\ref{eqn:T3}). Note that, although we
have managed to rewrite the above equations in a way close to the ones
for the isolated case, they contain all the relevant in-medium
corrections in a systematic way, i.e. correct Pauli-blocking and self
energy corrections. The numerical solution requires some mild
approximations that are however well understood in the context of the
isolated few-body problem.

\section{Results}
\begin{figure}[t]
\begin{minipage}{0.49\textwidth}
\epsfig{figure=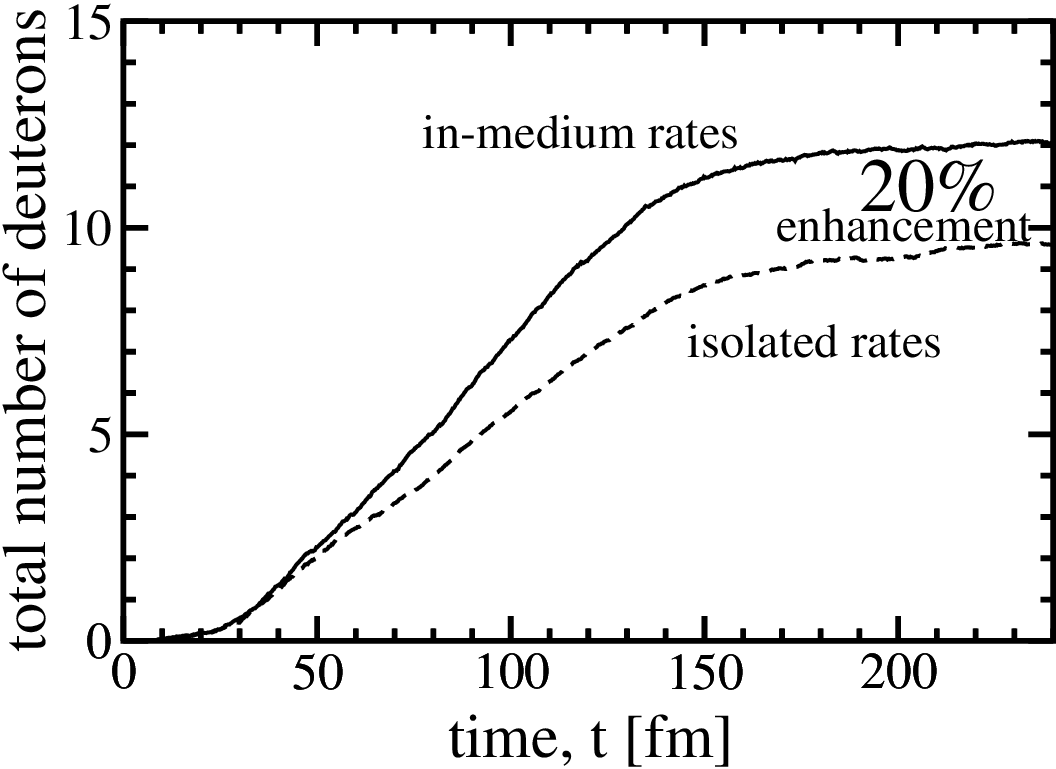,width=0.9\textwidth}
\caption{\label{fig:PLT} 
  BUU simulation of deuteron formation in the collision of
  $^{129}$Xe+$^{119}$Sn at 50 MeV/A. Use of in-medium rates lead to a
  20\% enhancement.}
\end{minipage}\hfill
\begin{minipage}{0.49\textwidth}
  \epsfig{figure=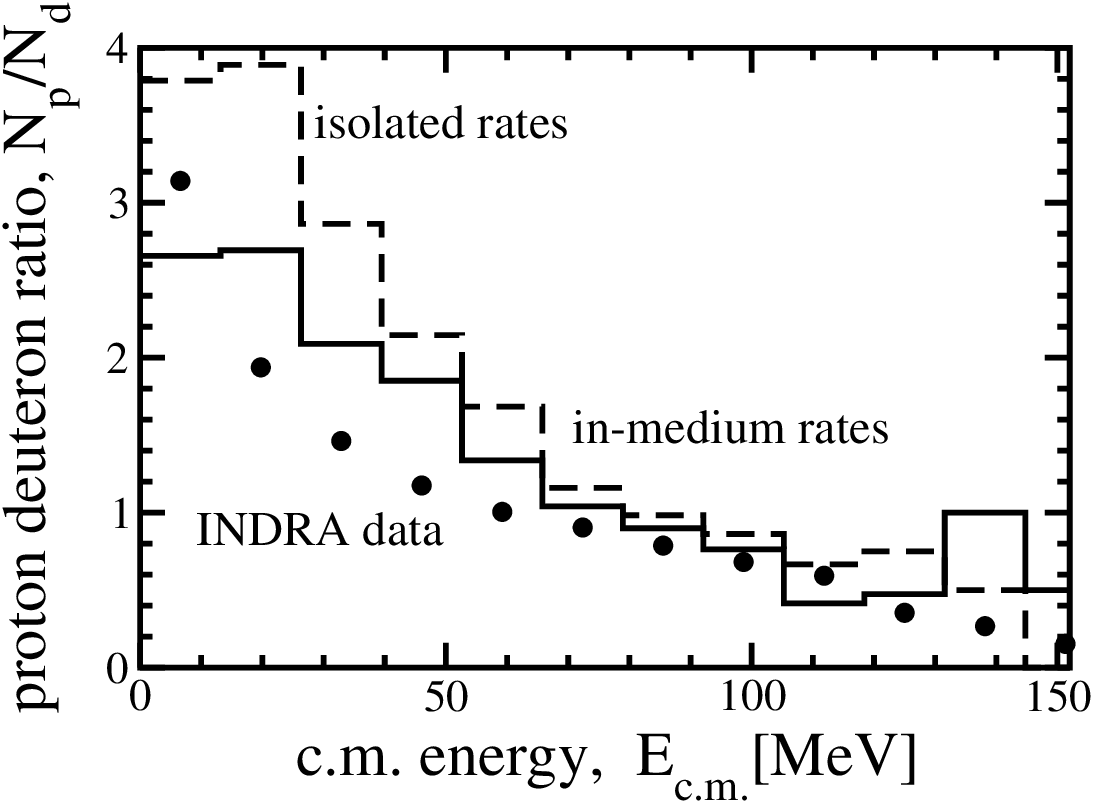,width=0.9\textwidth}
\caption{\label{fig:pd} 
  Ratio of proton to deuteron numbers as a function of c.m. energy. The
  experimental data are from the INDRA collaboration.}
\end{minipage}
\end{figure}
An experiment to explore the equation of state of nuclear matter is
heavy ion collisions at various energies. Here we focus on
intermediate to low scattering energies and compare results to an
experiment $^{129}$Xe+$^{119}$Sn at 50 MeV/A by the INDRA
collaboration~\cite{INDRA}. A microscopic approach to tackle the heavy
ion collision is given by the Boltzmann equation for different
particle distributions and solved via a Boltzmann Uehling Uhlenbeck
(BUU) simulation ~\cite{Kuhrts:2001zs}. The reaction rates
appearing in the collision integrals are {\em a priori} medium
dependent. 
\begin{figure}[b]
\begin{minipage}{0.49\textwidth}
\epsfig{figure=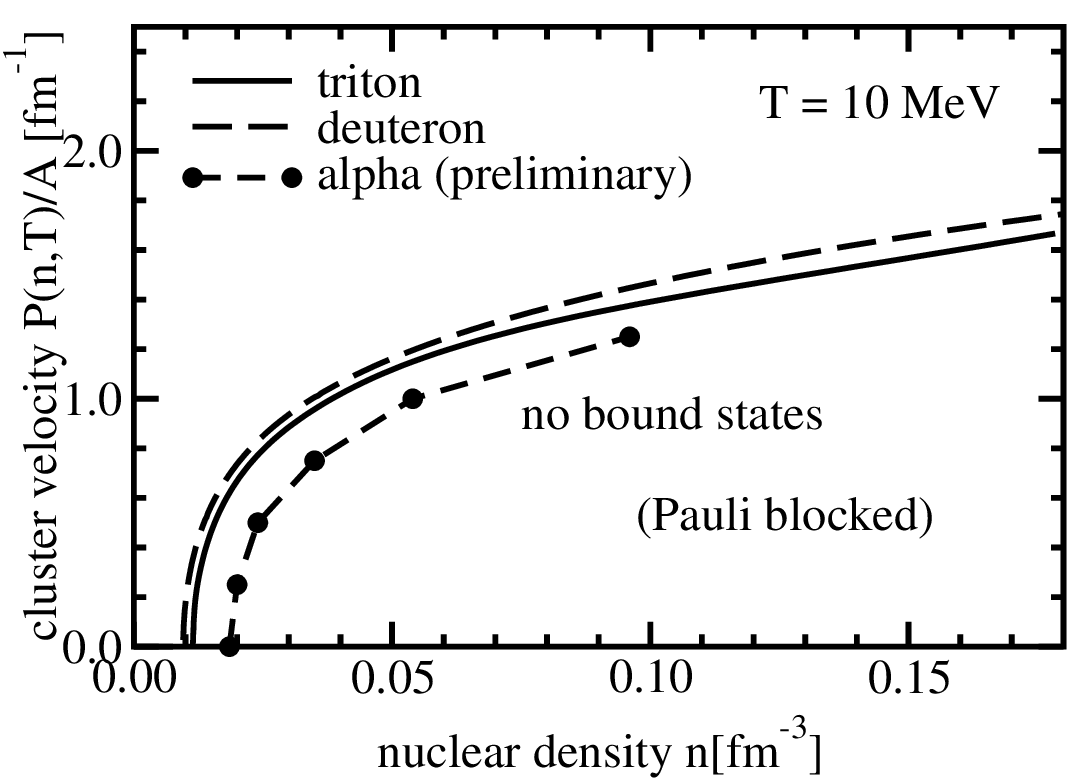,width=\textwidth}
\caption{\label{fig:amottV} 
  Due to the Mott effect clusters exist only above the respective
  curves. Below no bound states exist because of Pauli blocking.}
\end{minipage}\hfill
\begin{minipage}{0.49\textwidth}
\psfig{figure=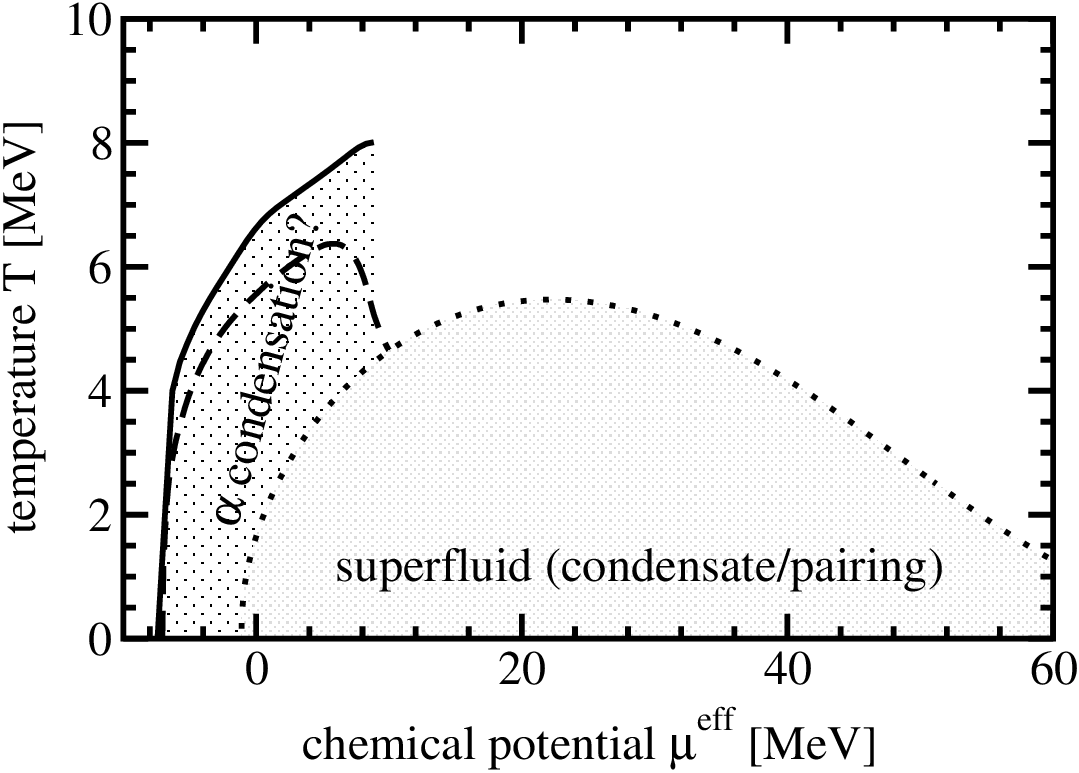,width=\textwidth}
\caption{\label{fig:Tc} 
  Part of the phase diagram with lines of critical temperatures.
  Solid line four-body AGS type calculation. Others earlier
  calculation~\cite{ alpha}}
\end{minipage}
\end{figure}
We use the in-medium AGS equations (\ref{eqn:T3}) that reproduce the
experimental data in the limit of an isolated three-body system.  For
details on the specific interaction model see
Ref.~\cite{Beyer:1996rx}. We investigate the influence of medium
dependent rates in the BUU simulation of the heavy ion collision as
compared to use of isolated rates.
Figure~\ref{fig:PLT} shows that the net effect (gain-minus-loss) of
deuteron production becomes larger for the use of in-medium rates
(solid) compared to using the isolated rates (dashed). The change is
significant, however, a comparison with experimental data is difficult
since deuterons may also be evaporating from larger clusters that has
not been taken into account in the present calculation so far. The
ratio of protons to deuterons may be better suited for a comparison to
experiments that is shown in Figure \ref{fig:pd}. The use of in-medium
rates (solid) lead to a shape closer to the experimental data (dots)
than the use of isolated rates (dashed). 
Within linear response theory for infinite nuclear matter
the use of in-medium rates leads to faster time scales for the
deuteron life time and the chemical relaxation time as has been shown
in detail in Refs.~\cite{Beyer:1997sf,Kuhrts:2000jz}. This faster time
scales should have consequences for the freeze out of fragments.


As does the nucleon, see eq.~(\ref{eqn:quasi}), the cluster changes
its mass. The binding energy changes as well until the Pauli blocking
is too strong for bound states to exist (Mott effect). This effect
depends on the momentum of the cluster. This important effect for
modeling of heavy ion collisions is shown in Figure \ref{fig:amottV}.

In Figure \ref{fig:Tc} part of the phase diagram of nuclear matter is
shown.  The condition for the onset of superfluidity for
$\ga$-particles is $B(T_c,\mu_c,P=0)=4\mu_c$, where $B$ is the binding
energy. The critical temperature found by solving the homogeneous AGS
equation for $\mu<0$ confirms the onset of $\alpha$ condensation even
at higher values (solid line) as given earlier (dashed, from
\cite{alpha}) which was based on a variational calculation using the
2+2 component of the $\alpha$ particle.  For $\mu>0$ the condition
$E=4\mu$ for the phase transition can also be fulfilled.  However, the
significance for a possible quartetting needs further investigation.

To conclude, strongly correlated Fermi systems such as nuclear/quark
matter provide an excellent field for few-body techniques.  

\begin{acknowledge}
  Work supported by the Deutsche Forschungsgemeinschaft.
\end{acknowledge}

\end{document}